\documentclass[twocolumn]{aastex61}
\usepackage{epsfig}
\usepackage{epstopdf}
\usepackage{gensymb}
\usepackage{lipsum}
\usepackage{subfigure}
\usepackage{multirow}

\submitjournal{AAS Journals}

\shorttitle{Thermal Emission Spectrum of HAT-P-7b}
\shortauthors{Mansfield et al.}

\begin{document}

\title{A HST/WFC3 Thermal Emission Spectrum of the Hot Jupiter HAT-P-7b}

\correspondingauthor{Megan Mansfield}
\email{meganmansfield@uchicago.edu}

\author{Megan Mansfield}
\affiliation{Department of Geophysical Sciences, University of Chicago, 5734 S. Ellis Avenue, Chicago, IL 60637, USA}

\author{Jacob L. Bean}
\affiliation{Department of Astronomy \& Astrophysics, University of Chicago, 5640 S. Ellis Avenue, Chicago, IL 60637, USA}

\author{Michael R. Line}
\affiliation{School of Earth and Space Exploration, Arizona State University, Tempe, AZ 85281, USA}

\author{Vivien Parmentier}
\affiliation{Aix Marseille Universit\'e, CNRS, LAM, Laboratoire d'Astrophysique de Marseille, Marseille, France}

\author{Laura Kreidberg}
\affiliation{Harvard-Smithsonian Center for Astrophysics, Harvard University, Cambridge, MA 02138, USA}

\author{Jean-Michel D\'esert}
\affiliation{Anton Pannekoek Institute for Astronomy, University of Amsterdam, 1090 GE Amsterdam, Netherlands}

\author{Jonathan J. Fortney}
\affiliation{Department of Astronomy and Astrophysics, University of California, Santa Cruz, CA 95064, USA}

\author{Kevin B. Stevenson}
\affiliation{Space Telescope Science Institute, Baltimore, MD 21218, USA}

\author{Jacob Arcangeli}
\affiliation{Anton Pannekoek Institute for Astronomy, University of Amsterdam, 1090 GE Amsterdam, Netherlands}

\author{Diana Dragomir}
\affiliation{Kavli Institute for Astrophysics and Space Research, Massachusetts Institute of Technology, Cambridge, MA 02139, USA}
\affiliation{Hubble Fellow}

\begin{abstract}

Secondary eclipse observations of several of the hottest hot Jupiters show featureless, blackbody-like spectra or molecular emission features, which are consistent with thermal inversions being present in those atmospheres. Theory predicts a transition between warmer atmospheres with thermal inversions and cooler atmospheres without inversions, but the exact transition point is unknown. In order to further investigate this issue, we observed two secondary eclipses of the hot Jupiter HAT-P-7b with the \textit{Hubble Space Telescope} (\textit{HST}) WFC3 instrument and combined these data with previous \textit{Spitzer} and \textit{Kepler} secondary eclipse observations. The \textit{HST} and \textit{Spitzer} data can be well fit by a blackbody with $T=2692 \pm 14$\,K, and the \textit{Kepler} data point constrains the geometric albedo to $A_{g}=0.077 \pm 0.006$. We modeled these data with a 3D GCM and 1D self-consistent forward models. The 1D models indicate that the atmosphere has a thermal inversion, weak heat redistribution, and water dissociation that limits the range of pressures probed. This result suggests that WFC3 observations of HAT-P-7b and possibly some other ultra-hot Jupiters appear blackbody-like because they probe a region near the tropopause where the atmospheric temperature changes slowly with pressure. Additionally, the 1D models constrain the atmospheric metallicity ($[\text{M/H}]=-0.87^{+0.38}_{-0.34}$) and the carbon-to-oxygen ratio (C/O\,$<1$ at 99\% confidence). The solar composition 3D GCM matches the \textit{Spitzer} data but generally underpredicts the flux in the WFC3 bandpass and cannot reproduce its featureless shape. This discrepancy could be explained by high atmospheric drag or nightside clouds, and may be better understood through further observation with the \textit{James Webb Space Telescope} (\textit{JWST}).

\end{abstract}


\section{Introduction}
\label{sec:intro}

Thermal emission measurements have revealed a wealth of information about the compositions and climates of hot Jupiter atmospheres. Composition determinations for these planets are important because they provide records of their formation and migration \citep{Venturini2016,Madhusudhan2017}. In particular, the core accretion model of planet formation predicts a trend of decreasing atmospheric metallicity with increasing planet mass for giant planets \citep{Fortney2013}. Additionally, the carbon-to-oxygen abundance ratio (C/O) provides information on the environment in which a planet forms and how the planet migrated to its current location \citep{Oberg2011,Madhusudhan2014,Mordasini2016,Espinoza2017,Ali2017}.

Beyond constraints on its composition, the thermal structure of a hot Jupiter provides information on its climate \citep{Showman2002,Burrows2006}. These highly irradiated, tidally locked planets are in a regime different from any observed in the Solar System, and so exhibit unique phenomenon, such as high temperatures, large day-night temperature differences, and high wind speeds \citep{Showman2002}. Additionally, molecules like TiO or VO can absorb incoming stellar flux and heat the upper atmosphere, creating a thermal inversion \citep{Hubeny2003}.

Hot Jupiter atmospheres are expected to show a variety of thermal structures ranging from warmer to cooler planets \citep{Fortney2008}. From an observational standpoint, planets with cooler dayside temperatures ($T \lessapprox2000$\,K), such as HD 189733b, HD 209458b, and WASP-43b, show molecular absorption bands in their emission spectra, indicating that they have temperature-pressure (T-P) profiles exhibiting decreasing temperature with increasing altitude \citep{Grillmair2008,Line2016,Kreidberg2014b,Stevenson2014c}. However, the emission spectra of hotter planets ($T \gtrapprox 2000$\,K) are inconsistent with such T-P profiles. \textit{HST}/WFC3 observations of WASP-12b and WASP-103b from 1.1-1.7 $\mu$m have revealed blackbody-like emission spectra, which could be indicative of isothermal T-P profiles \citep{Swain2013,Cartier2017}. In addition, the planets WASP-33b and WASP-121b show signs of molecular emission bands, which indicate a thermal inversion \citep{Haynes2015, Evans2017}. Complicating the picture of a possible smooth transition from non-inverted to inverted thermal structures is the result for Kepler-13Ab, which shows molecular absorption features indicative of a non-inverted profile, possibly because of cold trapping in its atmosphere \citep{Beatty2017}. WASP-12b also shows signs of molecular absorption in the \textit{Spitzer} bandpass, despite its blackbody-like spectrum between 1.1-1.7 $\mu$m \citep{Madhusudhan2011,Stevenson2014a}.

In this paper we present the secondary eclipse spectrum of HAT-P-7b observed with \textit{HST}/WFC3 in the 1.1-1.7~$\mu$m range. HAT-P-7b is a hot Jupiter with an intermediate dayside equilibrium temperature of 2600\,K (assuming zero albedo and dayside only recirculation). HAT-P-7b was previously observed with \textit{Spitzer} and found to have a thermal inversion \citep{Christiansen2010,Wong2016}. However, because \textit{Spitzer} can only observe exoplanet spectra in a few broadband regions, there are degeneracies with the molecular abundances that make it difficult to determine the exact thermal structure with these data alone \citep{Madhusudhan2010,Line2014}. Spectroscopy in general, and WFC3 measurements in particular, can remove these degeneracies by resolving molecular bands. We describe our \textit{HST} observations of HAT-P-7b in \S \ref{sec:observe}. In \S \ref{sec:analysis}, we describe our data analysis and results, and we summarize our findings in \S \ref{sec:discuss}.

\section{Observations}
\label{sec:observe}

We observed secondary eclipses of HAT-P-7b on 23 December 2016 and 4 January 2017 using the \textit{HST} WFC3 IR detector as part of program GO-14792. We used the G141 grism to observe the emission spectrum of HAT-P-7b between 1.1-1.7~$\mu$m. Each of the two visits consisted of five consecutive \textit{HST} orbits, in which HAT-P-7 was visible for approximately 50 minutes per orbit and occulted by the Earth for the remainder of each orbit. At the beginning of each orbit, we took a direct image of the target with the F126N narrow-band filter for wavelength calibration.

\begin{figure}
\centering
\includegraphics[width=\linewidth]{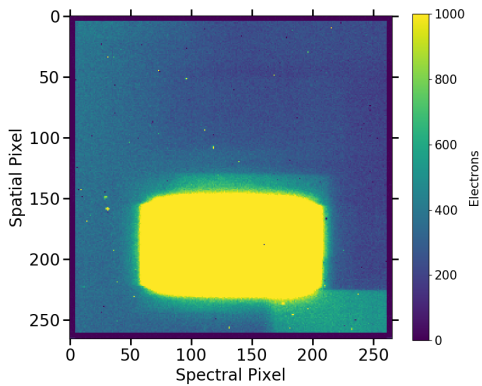}
\caption{\label{fig:scan} An example spatial scan. The spectrum of HAT-P-7 overlaps with a background star in the full image here, but our data reduction used the individual ramp samples, in which the two stars are well separated.}
\end{figure}

The observations were taken in spatial scan mode with the $256 \times 256$ subarray using the SPARS10, NSAMP=16 readout pattern, resulting in a total exposure time of 103.129\,s. We used a scan rate of 0.08 arcsec/second, which produced spectra extending approximately 80 pixels in the spatial direction and with peak pixel counts of about 35,000 electrons per pixel. We used bi-directional scans to maximize the duty cycle, which yielded 21 exposures per orbit and a duty cycle of 64\%. An example spatial scan is shown in Figure \ref{fig:scan}. Although the spectrum of a background star overlaps with that of HAT-P-7 in the full image, our data reduction used the individual ramp samples, in which the two stars are well separated.

We reduced the \textit{HST} data using the data reduction pipeline described in \citet{Kreidberg2014a}. All observation times were converted to BJD$_{\text{TDB}}$ \citep{Eastman2010}. We used an optimal extraction procedure instead of aperture extraction to extract the data \citep{Horne1986}. We tested several different aperture sizes to determine one that captured the full spectrum without capturing large areas of background. We masked cosmic rays so that optimal extraction could fit the point spread function of the data without being influenced by cosmic rays. A typical frame had one cosmic ray masked out. To subtract the background out of each frame, we visually inspected the images to find a clear background spot on the detector and subtracted the median of this background area from each pixel in the aperture. To determine the uncertainties on the measurements we added the photon noise, read noise, and median absolute deviation of the background in quadrature.

Following standard procedure for \textit{HST} WFC3 eclipse observations, we discarded the first orbit of each visit. We also discarded the first exposure from each orbit in the first visit and the first two exposures from each orbit in the second visit to improve the quality of the fit. The spectra were binned into 14 channels with a width of 9 pixels per channel, giving a resolution of $R=30-44$. We also created a broadband white light curve by summing the spectra over the entire wavelength range.

We fit both the white light curve and the spectroscopic light curves with a model that combined a secondary eclipse model \citep{Kreidberg2015a} and a systematics model based on \citet{Berta2012}. For the white light curve, the free parameters in the secondary eclipse model were the mid-eclipse time $T_{sec}$ and the planet-to-star flux ratio $F_{p}/F_{s}$. The orbital period, ratio of the semi-major axis to the stellar radius, inclination, planet-to-star radius ratio, and eccentricity were fixed to the values determined by \citet{Wong2016}, which were $P= 2.2047372$~days, $a/R_{s}= 4.03$, $i=82.2\degree$, $R_{p}/R_{s}=0.07809$, and $e=0.0016$, respectively.

We fit the instrument systematics with an equation of the form
\begin{equation}
M(t)=E(t)(cs+vt_{vis})(1-e^{-r_{1}t_{orb}-r_{2}})
\label{eq:systematics}
\end{equation}
where $M(t)$ is the modeled flux, $E(t)$ is the eclipse model, $c$ is a normalization constant, $s$ is a scaling factor to correct for an offset in normalization between scan directions \citep{McCullough2012}, $v$ is a visit-long linear slope, $t_{vis}$ is the time since the beginning of the visit, $r_{1}$ is the amplitude of an orbit-long exponential ramp, $r_{2}$ is the time constant of the orbit-long ramp, and $t_{orb}$ is the time since the beginning of the orbit. $T_{sec}$, $f_{p}/f_{s}$, $r_{1}$, and $r_{2}$ were fixed to the same values for both visits, while $c$, $v$, and $s$ varied between visits. The fit to the white light curve thus contained a total of 10 free parameters.

Previous studies of \textit{HST} WFC3 data have shown that adding a quadratic term to the visit-long trend in the model of the instrument systematics provides a better fit, primarily for very bright stars \citep{Stevenson2014b, Line2016}. We tested adding a quadratic term to the visit-long trend, but found Bayesian Information Criterion (BIC) values that were higher by about 8 on average for the quadratic model compared to the linear model, indicating that the linear model is preferred for this data set.

\begin{figure}
\centering
\includegraphics[width=\linewidth]{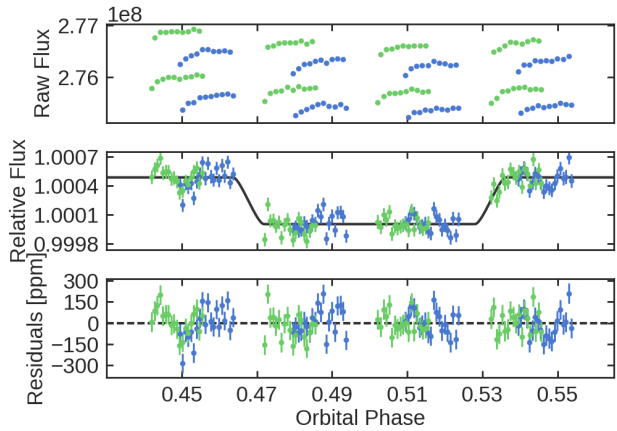}
\caption{\label{fig:whitelight} Raw flux of the secondary eclipse of HAT-P-7b (upper panel), best fit broadband white light curve (middle panel), and residuals to the fit (lower panel). Blue points are from the eclipse on 23 December 2016, and green points are from 4 January 2017. The offsets between the two sets of green and blue points in the upper panel are due to the difference between the forward and reverse scans. The fit has $\chi^{2}_{\nu}=2.24$ and an average residual of 90~ppm.}
\end{figure}

\begin{figure*}
\centering
\includegraphics[width=\linewidth]{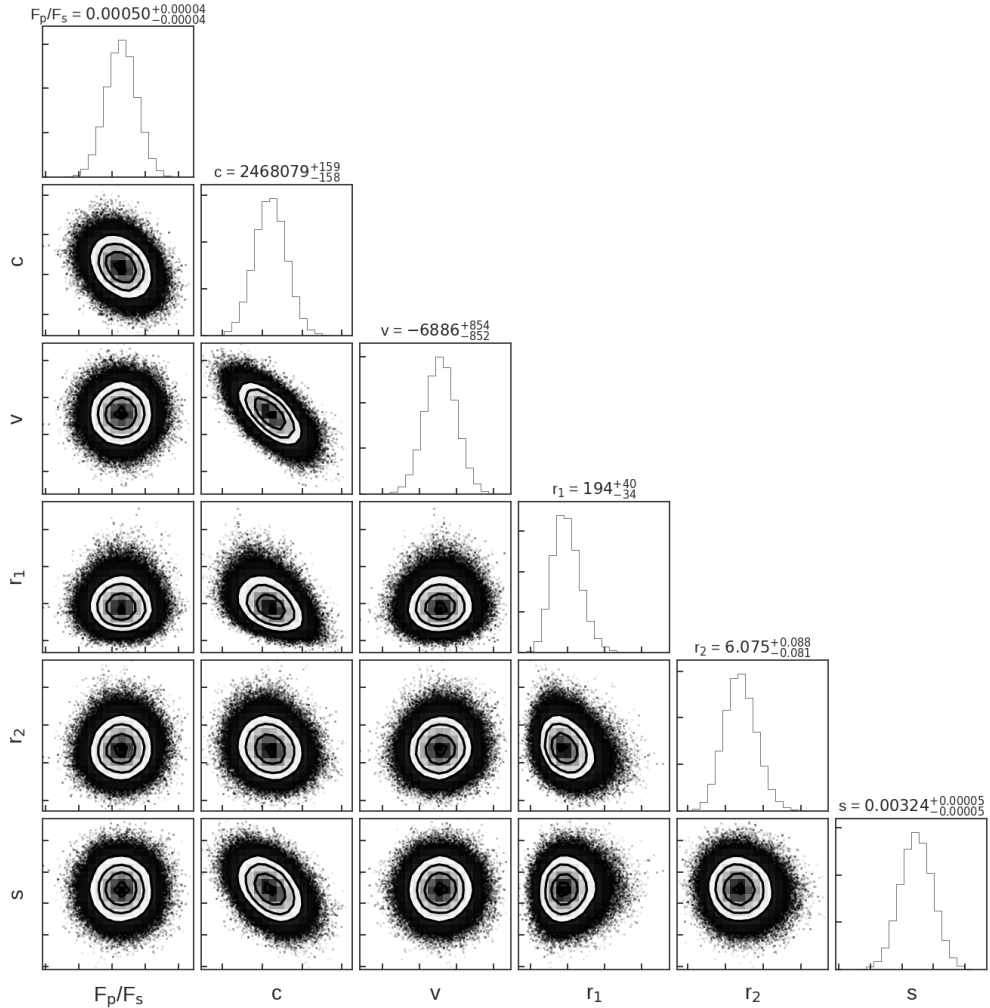}
\caption{\label{fig:pairs} Pairs plot showing the distributions of fit parameters for the MCMC fit to the 1.234 - 1.271~$\mu$m light curve. The off-diagonal panels show marginalized posterior probability for pairs of parameters, with 1, 2, and 3$\sigma$ intervals indicated with black contours. The grey shading is darker for higher probability density. The panels on the diagonal show marginalized posterior probability distributions for each parameter, and the dashed lines indicate the median values and 68\% confidence intervals. The planet-to-star flux ratio is not strongly correlated with any of the other fit parameters. For parameters that are allowed to vary between visits ($c$, $v$, and $s$), the distributions are for the eclipse observed on 23 December 2016.}
\end{figure*}

\begin{figure*}
\centering
\includegraphics[width=\linewidth]{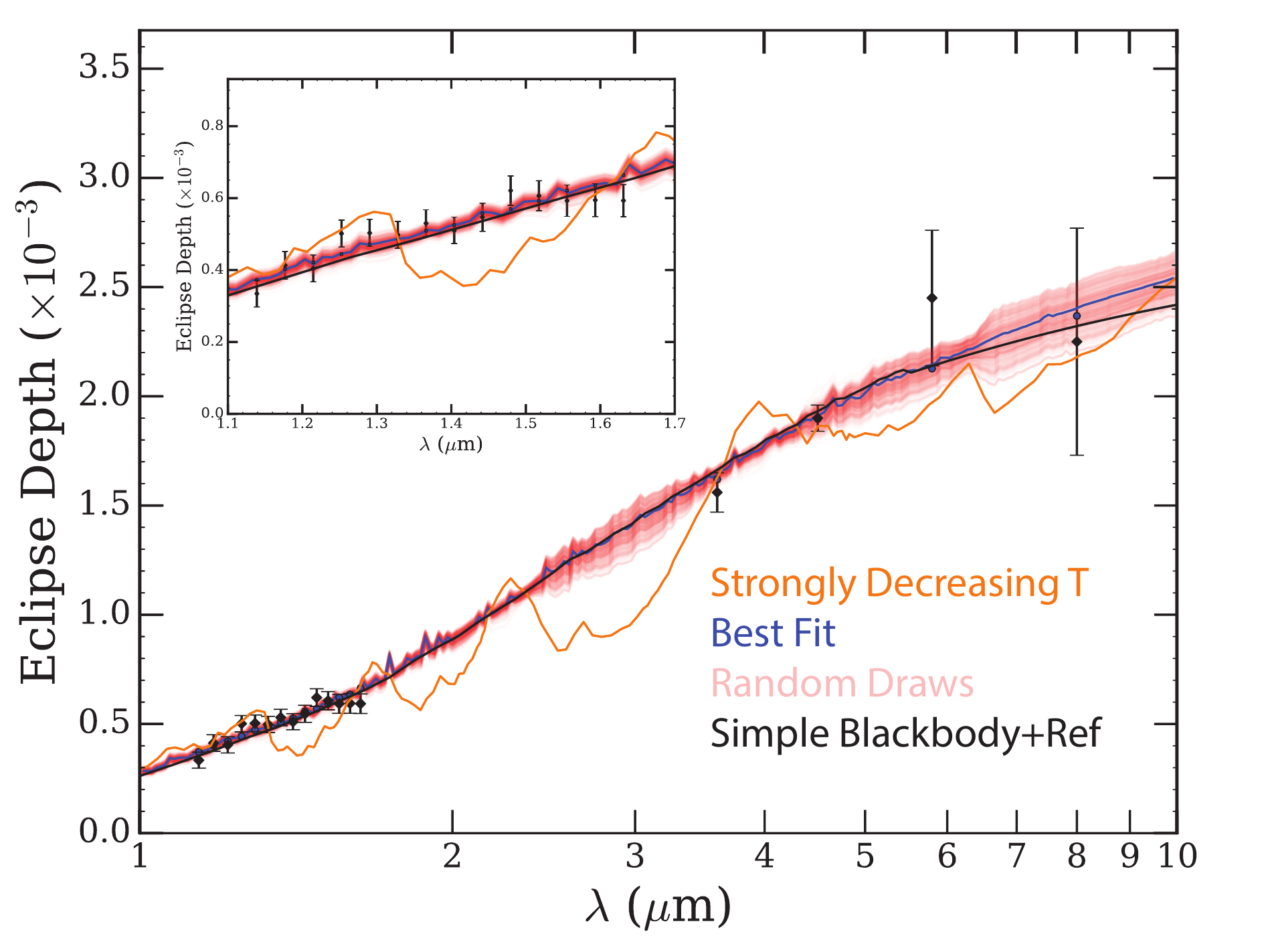}
\caption{\label{fig:fitspec} Secondary eclipse spectrum with a suite of theoretical models. Black points with $1\sigma$ error bars represent observations by WFC3 in this paper and by \textit{Spitzer} in \citet{Wong2016}. The insert shows the WFC3 data from 1.1-1.7~$\mu$m. The dark blue line represents the best-fitting 1D atmospheric model, as described in \S \ref{sec:1d} , and the surrounding red lines show 500 spectra randomly drawn from the posterior. Blue points outlined in black show the best-fitting 1D model binned to the data resolution. The black line shows a fit to a simple model of thermal and reflected light, as described in \S \ref{sec:thermal}. The orange line shows a model atmosphere with a monotonically decreasing temperature-pressure profile that provides a reasonable match to the \textit{Spitzer} data, calculated using the methods of \citet{Fortney2008}.}
\end{figure*}

We estimated the parameters with a Markov Chain Monte Carlo (MCMC) fit using the {\fontfamily{bch}\selectfont emcee} package for Python \citep{Foreman2013}. The best-fit white light curve, which is shown in Figure \ref{fig:whitelight}, had $\chi^{2}_{\nu}=2.24$ and an average residual of 90 ppm. The value of $T_{sec}$ determined from the fit to the white light curve was $2457757.68242 \pm 0.00097$~BJD$_{\text{TDB}}$. The spectroscopic light curves were fit with the same model as the white light curve, with the exception that $T_{sec}$ was fixed to the best-fit value from the fit to the white light curve. An example pairs plot for the MCMC fit to the 1.234 - 1.271~$\mu$m light curve is shown in Figure \ref{fig:pairs}. The spectroscopic light curves achieved photon-limited precision, with $\chi^{2}_{\nu}$ values between $0.89-1.25$. The final secondary eclipse spectrum, along with \textit{Spitzer} data from \citet{Wong2016}, is shown in Figure \ref{fig:fitspec}, and the planet-to-star flux ratio for each bandpass is listed in Table \ref{tab:data}.

\begin{table}
\centering
\caption{Secondary eclipse spectrum of HAT-P-7b.}
\label{tab:data}
\begin{tabular}{c | c}
\hline \hline
Wavelength Range ($\mu$m) & $f_{p}/f_{s}$ (\%) \\
\hline
1.120 - 1.158 & $0.0334 \pm 0.0037$ \\
1.158 - 1.196 & $0.0413 \pm 0.0038$ \\
1.196 - 1.234 & $0.0404 \pm 0.0037$ \\
1.234 - 1.271 & $0.0501 \pm 0.0037$ \\
1.271 - 1.309 & $0.0503 \pm 0.0038$ \\
1.309 - 1.347 & $0.0498 \pm 0.0037$ \\
1.347 - 1.385 & $0.0530 \pm 0.0037$ \\
1.385 - 1.423 & $0.0510 \pm 0.0037$ \\
1.423 - 1.461 & $0.0547 \pm 0.0039$ \\
1.461 - 1.499 & $0.0621 \pm 0.0041$ \\
1.499 - 1.536 & $0.0607 \pm 0.0042$ \\
1.536 - 1.574 & $0.0593 \pm 0.0044$ \\
1.574 - 1.612 & $0.0594 \pm 0.0046$ \\
1.612 - 1.650 & $0.0593 \pm 0.0045$ \\
\hline

\end{tabular}
\end{table}

\section{Analysis}
\label{sec:analysis}

The spectrum of HAT-P-7b is shaped like a blackbody and clearly rejects a model with a monotonically decreasing temperature with altitude, as can be seen in Figure \ref{fig:fitspec}. This could be due to a lack of near-infrared opacity sources like water in the atmosphere, an isothermal atmospheric structure, or a previously unrecognized grey opacity obscuring absorption or emission features. To understand why we see this blackbody-like spectrum, we used three different modeling approaches: a 3D GCM, 1D self-consistent forward models, and a simple model of blackbody thermal emission plus reflected stellar light. The data are well-fit by a blackbody model, so we use these models with varying amounts of complexity to explore the planetary physics and chemistry and put the blackbody-like spectrum in context, rather than using fit quality metrics to search for a single best-fit model.

\subsection{Fit to 3D GCM}

\begin{figure}
\centering
	\begin{subfigure}{}
	\includegraphics[width=\linewidth]{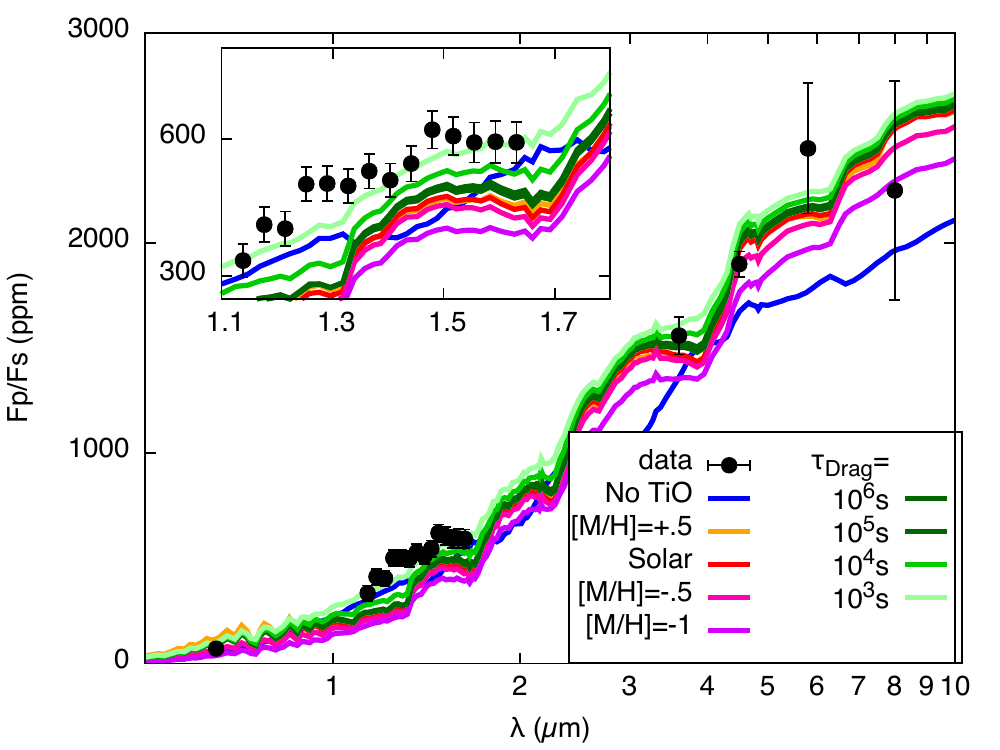}
	\end{subfigure}
	\begin{subfigure}{}
	\includegraphics[width=\linewidth]{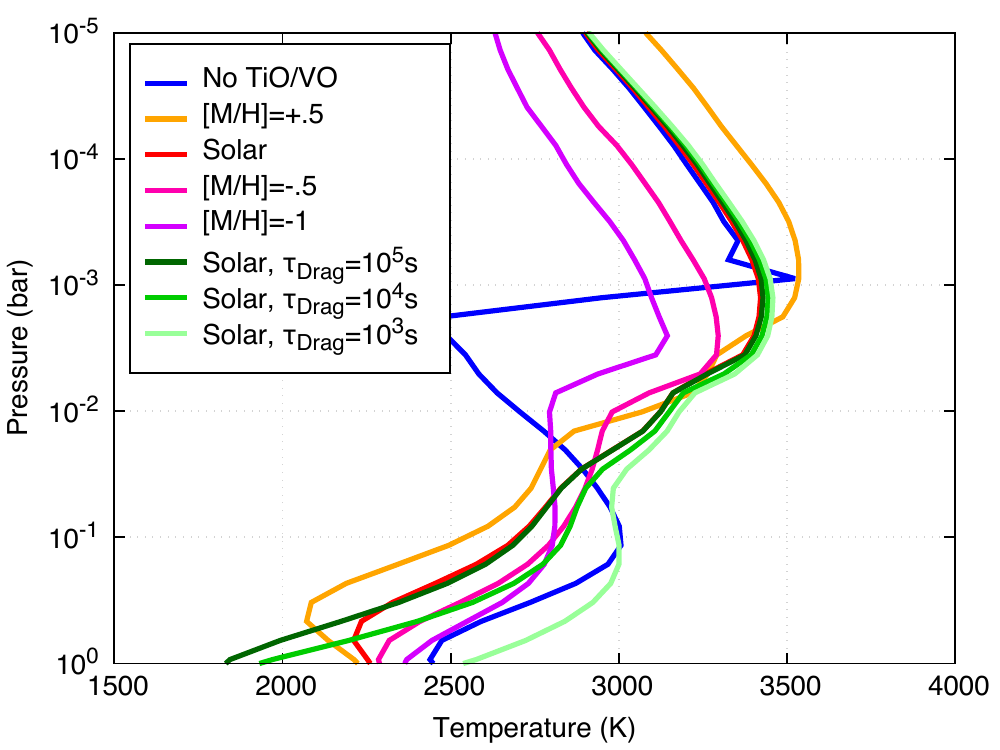}
	\end{subfigure}
\caption{\label{fig:Vivien} Planet-to-star flux ratio as a function of wavelength (top) and dayside temperature-pressure profiles (bottom) for each 3D GCM run. The red, blue, orange, pink, and magenta lines show models with solar composition, without TiO and VO, and with metallicities of $+0.5$, $-0.5$, and $-1.0$, respectively. The dark to light green lines show atmospheres with solar composition and decreasing drag timescales.}
\end{figure}

We first performed 3D GCM calculations to help interpret the spectrum and guide how we approached fitting the data with parameterized models. To model the three-dimensonal structure of HAT-P-7b, we used the SPARC/MITgcm clobal circulation model \citep{Showman2009}. Our setup is similar to the one used in \citet{Parmentier2016} but with planetary and stellar parameters chosen to match the HAT-P-7 system. The atmospheric opacities and mean molecular weight used in the calculations correspond to a solar composition atmosphere, a solar composition atmosphere depleted in TiO and VO, or a solar composition atmosphere with the abundance of every element apart from hydrogen and helium increased or decreased by the same amount. We assume local chemical equilibrium. Atmospheric drag of various possible origins (ohmic dissipation, hydrodynamic instabilities, etc.) is modeled as a Rayleigh drag present throughout the whole model and acting with a drag timescale $\tau_{\rm Drag}$. The model was run for 300 Earth days and all quantities were averaged over the last 100 days. Figure \ref{fig:Vivien} shows the planet-to-star flux ratio and T-P profile at the substellar point for each 3D model.

The best fitting 3D models contain both TiO/VO and H$^{-}$ opacities, and have a thermal inversion due to absorption by TiO/VO in the upper atmosphere. Models with lower metallicities have a deeper photosphere leading to more efficient heat redistribution and thus lower fluxes in the WFC3 bandpass. The low-metallicity models would therefore require even more drag than in the solar metallicity case to match the observations. Although we include H$^{-}$ opacity, it does not contribute significantly to the atmospheric opacity. However, water dissociation has a large impact on the observed spectrum. Water dissociation limits the range of pressures probed, which limits our observations to the part of the atmosphere near the tropopause and produces a blackbody-like spectrum. If water were not dissociating in the upper atmosphere, our observations would probe a region of the atmosphere which extended above the tropopause and we would see emission features.

Models containing TiO/VO with varying drag timescales produce similar quality fits to the \textit{Spitzer} data points, but the WFC3 points can only be fit well with a short drag timescale of $\tau_{\rm Drag}=10^{3}$\,s. Models with less drag (i.e. more redistribution) do not produce hot enough daysides to match the WFC3 data. Preliminary calculations of   $\tau_{\rm Drag}$, which scales with the square of the planetary magnetic field, suggest that such a small $\tau_{\rm Drag}$ is not unrealistic for such hot planets \citep{Perna2010}. However, the models with TiO/VO opacities also appear to have small emission features, which do not match the blackbody-like shape of the WFC3 data. One other possible explanation for the warmer dayside is nightside clouds, which would increase the greenhouse effect without increasing the planetary albedo \citep{Stevenson2014c,Kataria2015,Parmentier2016,Stevenson2017}. Nightside clouds on HAT-P-7b were also suggested by \citet{Armstrong2016} as an explanation for time variability in the brightness offset of its phase curve. It remains to be seen whether Lorentz forces or nightside clouds are the real explanation for why the nominal GCM underpredicts the WFC3 data.

\subsection{Fit to 1D grid models}
\label{sec:1d}

We also retrieve compositional and thermal information by fitting the spectrum with a grid of self-consistent 1D models.  We choose this self-consistent grid-based approach as spectra with little to no spectral features tend to drive classic retrievals \citep[e.g.,][]{Line2014} towards unphysical regions of parameter space.

The 1D models self-consistently solve for the radiative-convective-thermochemical equillbirium atmosphere solution.  For the radiative transfer, we use the \citet{Toon1989} two stream source function technique to solve for planetary thermal fluxes at each atmospheric level.  Incident stellar flux at the top of the atmosphere \citep[from a PHOENIX model --][]{Husser2013} is treated as a simple exponential attenuation at an average cosine incident angle of 1/$\sqrt{3}$.   The Newton-Raphson iteration technique is used to determine the layer temperatures that ensure zero net flux divergence across each model layer.  Opacities for H$_{2}$O, CH$_{4}$, CO, CO$_{2}$, NH$_{3}$, HCN, C$_{2}$H$_{2}$, H$_{2}$S, Na, K, TiO, VO, FeH, H$_{2}$-H$_{2}$/He CIA \citep{Lupu2014}, and H$^{-}$ bound-free and free-free \citep{John1988,Bell1987} are treated within the ``on-the-fly'' correlated-K framework \citep[e.g.,][]{Lacis1991,Amundsen2016}. H$_{2}$ and He Rayleigh scattering are added in as a continuum absorber.  Molecular abundances are computed using the NASA CEA Gibbs free energy minimization code \citep{Gordon1994} given the \citet{Lodders2009} elemental abundances.   

We reduce our parameterization of the 1D atmospheres to 3 parameters: day-night redistribution (f), which scales the incident stellar flux,  a metallicity, [M/H], which scales the \citet{Lodders2009} elemental abundances in the CEA routine, and a carbon-to-oxygen ratio (C/O).  A grid of 1D models are computed along this 3-vector parameter set with the f ranging between 1 and 4 in steps of 0.25, [M/H] from -1.5 to 2.5 dex in steps of 0.5 dex, and C/O between 0.1 and 2 on a non-uniform grid that more finely samples C/O values near 1 \citep[e.g.,][]{Molliere2015}.  The grid spacing is chosen to be fine enough that interpolation errors are negligible, and the grid range is chosen to be broad enough to capture a physically sensible range of values. In all of these models we assume that the relative abundances of all elements except C and O remain constant.

To fit the 3-parameter model grid spectra (at an R=100) to the data,  we use the {\fontfamily{bch}\selectfont emcee} package \citep{Foreman2013} combined with the Python {\fontfamily{bch}\selectfont griddata} N-dimensional interpolation routine. Only \textit{HST} and \textit{Spitzer} data were used in the fit. The best-fitting spectrum is shown in Figure \ref{fig:fitspec}. We include uniform prior ranges on $f$, [M/H], and log(C/O) between $1.75-2.66$, -$1.5-2.5$, and -$1.0-0.3$, respectively. We explored values of $f>2.66$ but decided to exclude them due to energy balance arguments that values larger than 2.66 violate energy conservation \citep[e.g.,][]{Cowan2011}.

\begin{figure}
\includegraphics[width=\linewidth]{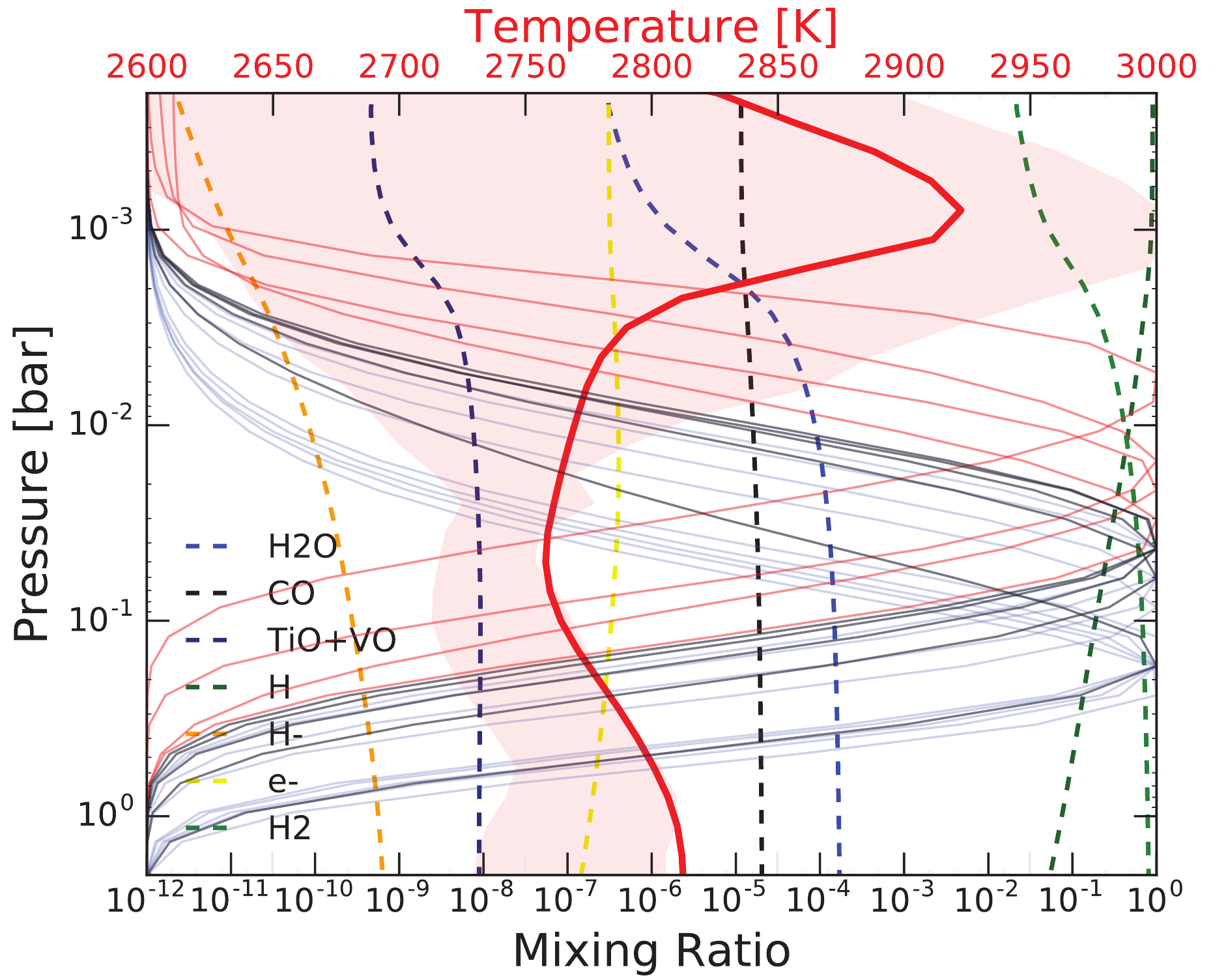}
\caption{\label{fig:tpgrid} Temperature-pressure profile for the best-fit 1D interpolated spectrum (bold red line) and 1$\sigma$ error (red shaded area). Temperatures are shown on the top x-axis. The model has a thermal inversion. Thin, solid lines indicate the contribution functions for data points in the \textit{Spitzer} bandpass (red), and the WFC3 bandpass (blue), with darker blue lines showing points inside the water band from 1.33-1.48 $\mu$m. The measurements probe between about 0.2 bar and 2 mbar, where the T-P profile changes temperature gradually. The dashed curves show thermo-chemical equilibrium mixing ratios for important absorbing molecules, computed along the self-consistent T-P profile shown in bold red. Mixing ratios are shown on the bottom x-axis.}
\end{figure}

Like the 3D GCM, the 1D grid models show a thermal inversion due to TiO/VO absorption heating the upper atmosphere. Figure \ref{fig:tpgrid} shows the T-P profile for the best-fitting grid model in bold red, with 1$\sigma$ errors representing the spread in the self-consistent T-P profiles that fall within the posterior shown by the red shaded area. Thin red and blue lines show the contribution functions for \textit{Spitzer} and WFC3 data points, respectively, with dark blue lines showing data in the water band from 1.33-1.48 $\mu$m. The contribution functions suggest that the data are primarily probing a region of the atmosphere near the tropopause where the temperature profile switches from non-inverted to inverted.

\begin{figure}
\includegraphics[width=\linewidth]{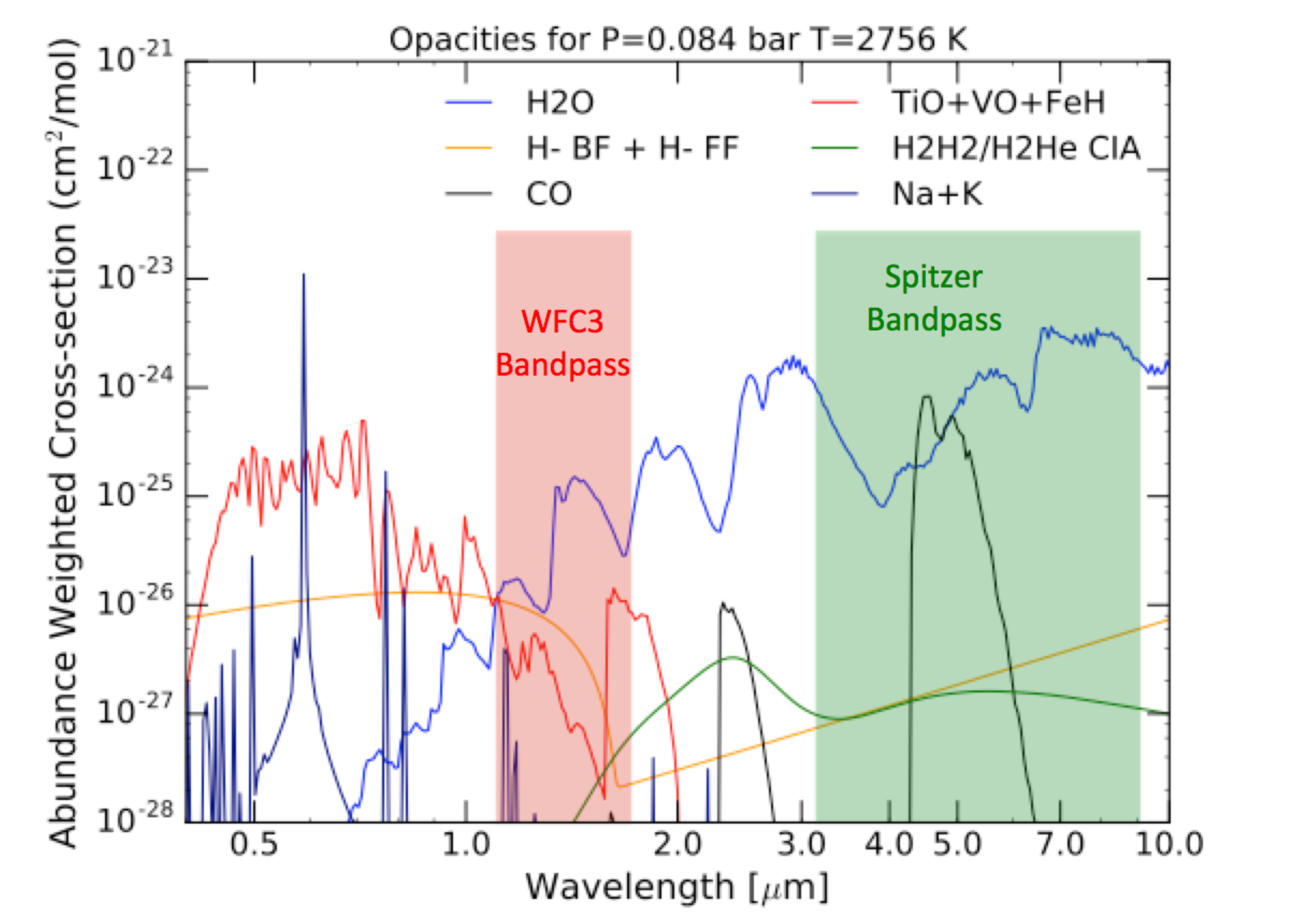}
\caption{\label{fig:opacity} Opacities of molecular water (blue), H$^{-}$ (orange), CO (black), TiO+VO+FeH (red), Na+K (dark blue), and hydrogen/hydrogen and hydrogen/helium collision-induced absorption (green) at a pressure of 0.084 bar in the best-fitting 1D model. The red shaded area indicates the WFC3 bandpass, and the green shaded area indicates the \textit{Spitzer} bandpass.}
\end{figure}

The dashed lines on Figure \ref{fig:tpgrid} show the thermo-chemical equilibrium mixing ratios for a set of important absorbing molecules, computed along the self-consistent T-P profile for the best fit. The water abundance begins to decrease rapidly around 2 mbar because water begins to dissociate in the hot upper atmosphere. This suggests that the WFC3 and \textit{Spitzer} observations probe the part of the atmosphere just below the tropopause because water dissociation limits the pressure range observed in that bandpass. If there were no water dissociation in the atmosphere, the \textit{Spitzer} observations would extend up to higher pressures in the atmosphere and we would observe an emission feature. However, water is still an important source of molecular opacity in the WFC3 bandpass, as shown in Figure \ref{fig:opacity}. This figure shows opacities at a pressure of 0.084 bar, near the part of the atmosphere sampled by the WFC3 bandpass outside the water band. The opacity from molecular water dominates over all other opacities at this pressure, including H$^{-}$ opacity, in contrast to what is found for WASP-18b in \citet{Arcangeli2018}.

\begin{figure}
\includegraphics[width=\linewidth]{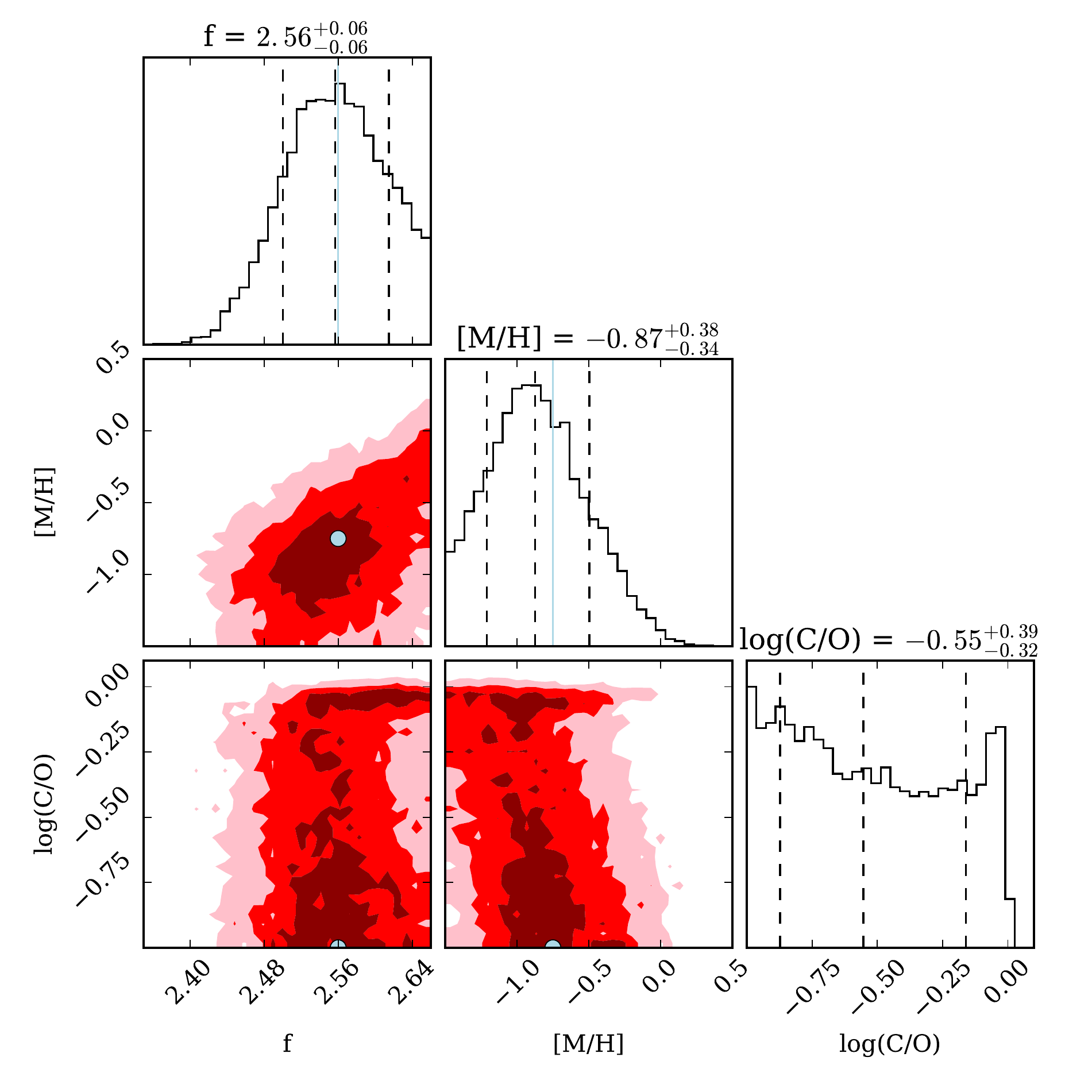}
\caption{\label{fig:mikepairs} Pairs plot for the 1D grid models, showing the heat redistribution ($f$), metallicity, and C/O ratio. The off-diagonal panels show marginalized posterior probability for pairs of parameters, with 1, 2, and 3$\sigma$ intervals indicated with dark red, red, and light red shading. The panels on the diagonal show marginalized posterior probability distributions for each parameter, and the dashed lines indicate the median values and 68\% confidence intervals. Blue lines and points indicate values for the best-fit interpolated spectrum.}
\end{figure}

Figure \ref{fig:mikepairs} shows a pairs plot for the 1D grid model fits. The key ingredient of this retrieval is the assumption of radiative-convective-thermochemical equilibrium. The factor $f$ measures heat redistribution, where $f=1$ indicates full redistribution, $f=2$ indicates redistribution over the dayside only, $f=2.66$ corresponds to zero heat redistribution and is the maximum possible value \citep{Cowan2011}. The best-fitting model has $f=2.56\pm0.06$, indicating that HAT-P-7b has relatively weak heat redistribution. This high value of $f$ likely suggests that the thermal emission we observe is being dominated by a localized ``hot spot'' on the planetary dayside.

\begin{figure*}
\centering
	\begin{subfigure}{}
	\includegraphics[width=\linewidth]{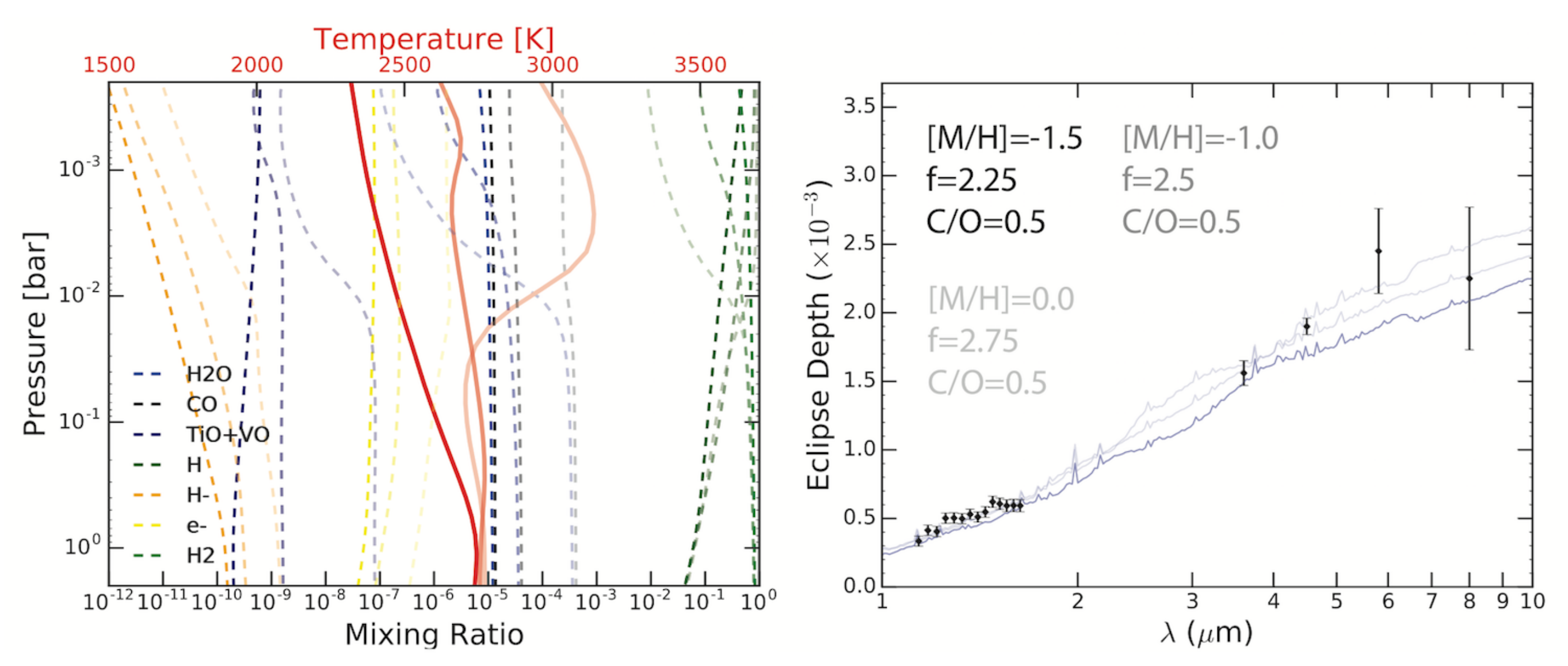}
	\end{subfigure}
	\begin{subfigure}{}
	\includegraphics[width=\linewidth]{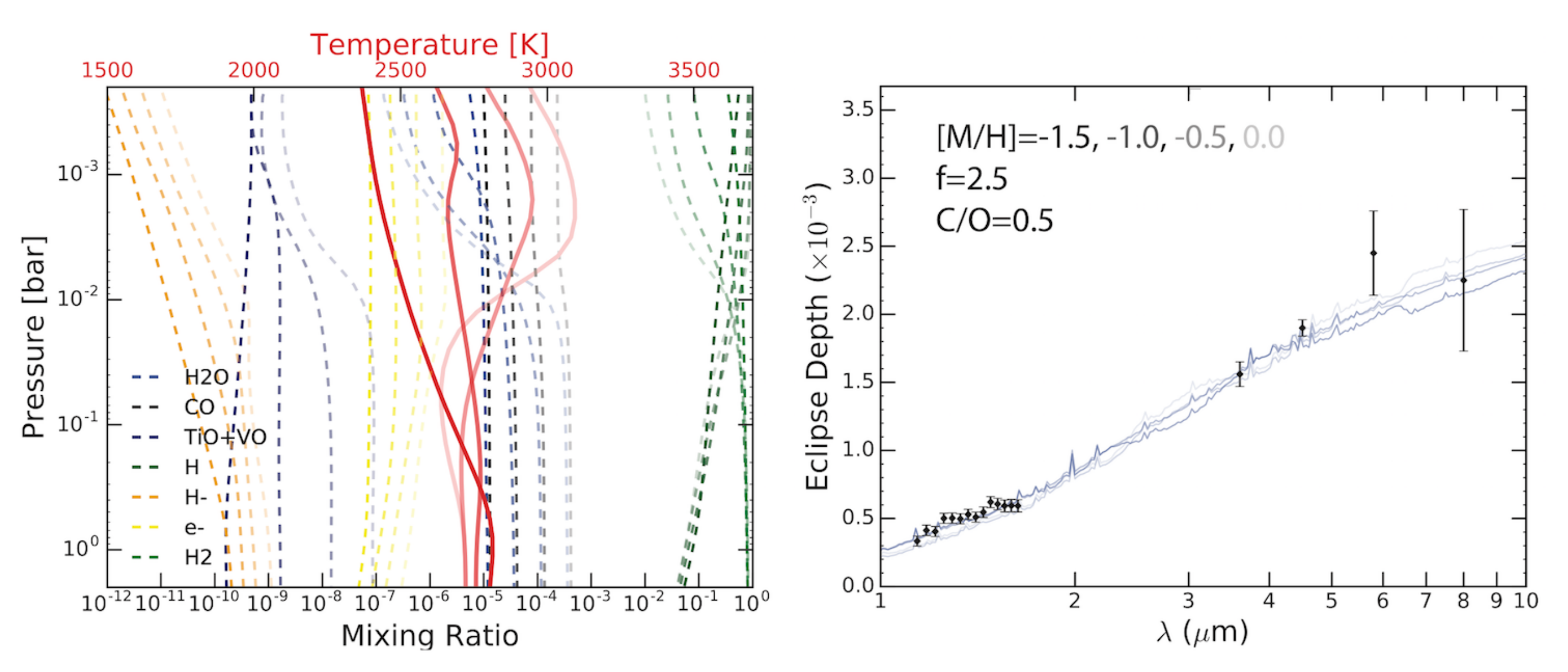}
	\end{subfigure}
\caption{\label{fig:diagnostics} Changes in the emission spectra and T-P profiles of the 1D models as a function of the metallicity ([M/H]) and heat redistribution ($f$). All of the T-P profiles and mixing ratios in the left panels are set up in the same manner as in Figure \ref{fig:tpgrid}, with the shades of each line corresponding to the dark, middle, and light spectra shown in the right panels. The top panels show that increasing both [M/H] and $f$ simultaneously produces the same temperature in the lower atmosphere. The bottom panels show how at higher [M/H] the inversion gets stronger and emission features begin to appear. The emission features are most apparent at high metallicities between $2-4$ $\mu$m.}
\end{figure*}

Although $f$ is highly correlated with the metallicity, the grid models produce a strong constraint on $f$. Changing the metallicity affects the value of $f$ because it changes the observed T-P profile. At higher metallicities, the abundances of TiO/VO and H$_{2}$O are both higher, but the abundances of TiO/VO increase faster than the abundance of H$_{2}$O. The higher TiO/VO abundances warm the upper atmosphere due to increased optical absorption, but the deeper atmosphere probed by WFC3 is cooler in order to maintain radiative equilibrium, as shown in Figure \ref{fig:diagnostics}. The cooling of the deeper atmosphere is then compensated for by increasing $f$, which warms the dayside by redistributing less of the heat to the nightside. This tradeoff can not continue indefinitely, however, because at very high metallicities the spectrum will begin to show emission features due to a stronger thermal inversion (Figure \ref{fig:diagnostics}). Similarly, at very low metallicities the spectrum will begin to show absorption features due to a monotonically decreasing T-P profile.  These changes in the metallicity are degenerate with changes in the relative abundances of Ti and O, but our model assumes that the ratio of Ti to O remains constant as the metallicity changes.

The models can also constrain the C/O because a higher C/O would decrease the abundances of TiO and VO in the atmosphere, which would weaken the thermal inversion. Therefore, the 1D models constrain the C/O\,$<1$ at 99\% confidence. This continues the trend of planets with C/O\,$<1$, with no planets having exotic high C/O values \citep{Benneke2015}. Decreasing the C/O, however, does not impact the spectrum shape or T-P profile because the TiO/VO opacity is only weakly dependent in this part of parameter space. Figure \ref{fig:mikepairs} demonstrates that the grid models only produce an upper limit on the C/O.

\subsection{Simple Two-Parameter Eclipse Model}
\label{sec:thermal}

Although the previous models only considered thermal emission from the planet, it has been suggested that reflected starlight may contribute significantly to the flux observed by \textit{HST} at near-infrared wavelengths \citep{Schwartz2015,Keating2017}. Reflected light can dominate over thermal emission in the near infrared for hot Jupiters with high albedos. \citet{Keating2017} found that reflected light may contribute significantly to the planetary flux at secondary eclipse for WASP-43b in the WFC3 bandpass. In order to determine whether reflected light is important in the WFC3 bandpass for HAT-P-7b, we performed a simple fit to a model of the combined reflected light and thermal emission for a planet whose emission can be described as a blackbody and that has a constant geometric albedo across all wavelengths. For such a planet, the flux from the planet can be described by
\begin{equation}
\frac{F_{p}}{F_{s}}=A_{g}\left(\frac{R_{p}}{a}\right)^{2}+\frac{B_{\lambda}(T_{d})}{F_{s,ph}}\left(\frac{R_{p}}{R_{s}}\right)^{2}
\end{equation}
where $\frac{F_{p}}{F_{s}}$ is the planet-to-star flux ratio, $A_{g}$ is the geometric albedo, $R_{p}$ is the planet radius, $R_{s}$ is the stellar radius, $a$ is the distance between the planet and star, $B_{\lambda}(T_{d})$ is the blackbody flux at the planetary dayside temperature $T_{d}$, and $F_{s,ph}$ is the stellar flux in each bandpass \citep{Keating2017}. $F_{s,ph}$ was determined by interpolating between Phoenix models \citep{Husser2013} using the Python package {\fontfamily{bch}\selectfont pysynphot} to obtain a model with $T=6441$\,K, $\log(g)=4.02$\,cm~s$^{-2}$, and [Fe/H]\,$=0.15$, which are the stellar parameters for HAT-P-7 \citep{Torres2012}. For this model, $B_{\lambda}(T_{d})$ and $F_{s,ph}$ were both integrated over the bandpasses of the data.

We fit this model to a combined data set including our WFC3 data and the \textit{Spitzer} and \textit{Kepler} data from \citet{Wong2016} using the {\fontfamily{bch}\selectfont emcee} Python package \citep{Foreman2013}. The free parameters in this fit were the geometric albedo $A_{g}$ and the planetary dayside temperature $T_{d}$. This fit is shown by the green line in Figure \ref{fig:fitspec}. We found HAT-P-7b to have a geometric albedo of $A_{g}=0.077 \pm 0.006$ and a dayside temperature of $T_{d}=2654 \pm 17$\,K. The albedo and dayside temperature are correlated, but well constrained by the combined data set because the infrared data, and the \textit{Spitzer} data in particular, are primarily sensitive to the temperature, while the \textit{Kepler} data add sensitivity to the reflected light. This estimate of the albedo is fairly consistent with previous estimates of HAT-P-7b's albedo \citep{Christiansen2010,Morris2013,Wong2016}.

Because the geometric albedo is a function of wavelength, it could be different in the \textit{Kepler} and WFC3 bandpasses. To ensure that fitting with a single geometric albedo was valid, we also performed a fit to the WFC3 and \textit{Spitzer} data including only thermal emission. This fit had a best-fit temperature of $T_{d}=2692 \pm 14$\,K We found through an F test that including an albedo parameter does not significantly improve the fit at these wavelengths longer than $1$~$\mu$m. Therefore, the WFC3 and \textit{Spitzer} data can be fit well by a single-temperature blackbody, and the albedo is constrained primarily by the planet-to-star flux in the \textit{Kepler} bandpass. Contrary to what \citet{Keating2017} found for WASP-43b, the contribution of reflected light in the WFC3 bandpass is insignificant for HAT-P-7b.

Although we found that HAT-P-7b has a low albedo, \citet{Sudarsky2000} suggest that very hot giant planets with effective temperatures over 1500\,K should have high albedos, as they should have silicate clouds forming high in their atmospheres. For example, MgSiO$_{2}$ should condense at a pressure of about 0.3~bar \citep{Sudarsky2000}. However, both our 3D and 1D models of HAT-P-7b predict the presence of TiO/VO high in its atmosphere, at pressures around 0.1-0.001~bar. The presence of such strong optical absorbers above the hypothesized silicate cloud deck could explain the low albedo that we observe.

Assuming that the planetary Bond albedo is equal to its geometric albedo, we can use the simple thermal and reflected light fit to calculate the heat redistribution across the surface of the planet. The heat redistribution is given by the equation
\begin{equation}
T_{d}=T_{0}(1-A_{B})^{1/4}\left(\frac{2}{3}-\frac{5}{12}\epsilon\right)^{1/4}
\end{equation}
where $T_{0}=T_{s}\sqrt{\frac{R_{s}}{a}}$ is the irradiation temperature, $T_{s}$ is the stellar temperature, $A_{B}$ is the Bond albedo, and $\epsilon$ is the redistribution efficiency \citep{Schwartz2015}. $\epsilon$ is the inverse of the parameter $f$ used in our 1D modeling, so smaller values of $\epsilon$ indicate less efficient heat redistribution. Using this equation, we find that the redistribution is $\epsilon=0.38 \pm 0.11$, which indicates that HAT-P-7b has very inefficient heat redistribution.

\section{Discussion}
\label{sec:discuss}

\begin{figure}
\includegraphics[width=\linewidth]{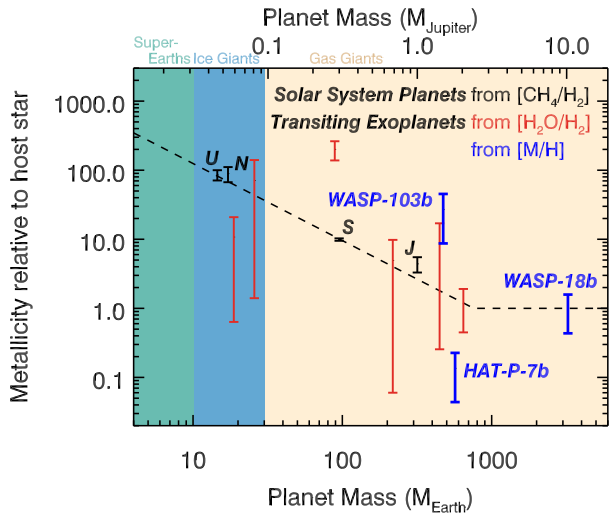}
\caption{\label{fig:massmetal} Trend in atmospheric metallicity vs. mass for solar system planets \citep[black points,][]{Wong2004,Fletcher2009,Karkoschka2011,Sromovsky2011} and exoplanets with visible water features (red points, see Table \ref{tab:metallicity}). Blue points show results for planets without clear molecular detections, for which the metallicities have been determined using the new self-consistent modeling developed in this paper and in \citet{Arcangeli2018}. The black dotted line is a fit to the values for the solar system planets, but plateauing at 1 once the planet metallicity equals the stellar metallicity.}
\end{figure}

1D grid modeling of the spectrum of HAT-P-7b suggests that the atmosphere contains a thermal inversion, and constrains the metallicity ($[\text{M/H}]=-0.87^{+0.38}_{-0.34}$) and carbon-to-oxygen ratio (C/O\,$<1$ at 99\% confidence). The new self-consistent 1D model developed in this paper, which assumes thermochemical and radiative-convective equilibrium, allows measurement of the atmospheric metallicity even though no molecular features are directly observed because of the feedback of metallicity onto the strength of the inversion. Figure \ref{fig:massmetal} shows the atmospheric metallicity as a function of planet mass for solar system planets \citep{Wong2004,Fletcher2009,Karkoschka2011,Sromovsky2011} and exoplanets. The exoplanet measurements plotted in Figure \ref{fig:massmetal} are based on either water abundances (red points) or the new self-consistent modeling developed in this paper (blue points). The exoplanet data plotted in this figure, and their sources, are listed in Table \ref{tab:metallicity}. The inferred metallicity of HAT-P-7b is well below the metallicity predicted by the trend in Figure \ref{fig:massmetal}. The low metallicity of HAT-P-7b could just be due to the expected intrinsic scatter around this trend \citep{Fortney2013}. Alternatively, \citet{Madhusudhan2014} predicted that sub-solar oxygen and carbon abundances could indicate a planet that formed farther away from its star and then underwent disk-free migration. They also predicted that planets which formed in this manner would have super-solar C/O ratios. Our upper limit on the C/O for HAT-P-7b does not definitively reveal whether its C/O ratio is super- or sub-solar, so more observations will be necessary to determine whether this theory can explain its low metallicity. This new modeling has the potential to reveal the compositions of other planets with blackbody-like spectra, such as WASP-12b \citep{Swain2013}. 

\begin{table*}
\centering
\caption{Exoplanet mass and metallicity data plotted in Figure \ref{fig:massmetal}}
\label{tab:metallicity}
\begin{tabular}{>{\centering \arraybackslash}p{1.9 cm} | >{\centering \arraybackslash}p{1.8 cm} | >{\centering \arraybackslash}p{2 cm} | >{\centering \arraybackslash}p{2.7 cm} | >{\centering \arraybackslash}p{2.6 cm} | >{\centering \arraybackslash}p{2.6 cm} | >{\centering \arraybackslash}p{2 cm}}
\hline \hline
Name & Planet Mass (M$_{\text{jup}})$ & Source & Planet Metallicity ($\times$ solar, $1\sigma$ range) & Source &  Stellar Metallicity ([M/H] or [Fe/H]) & Source \\
\hline
HAT-P-7b$^{\text{a}}$ & 1.78 & \citet{Pal2008} & $0.06-0.3$ & this work & $0.15 \pm 0.08$ & \citet{Torres2012} \\
\hline
HAT-P-11b$^{\text{b}}$ & 0.081 & \citet{Bakos2010} & $3.0-300$ & \citet{Fraine2014} & $0.33 \pm 0.07$ & \citet{Torres2012} \\
\hline
HAT-P-26b$^{\text{b}}$ & 0.059 & \citet{Hartman2011} & $0.8-26.3$ & \citet{Wakeford2017a} & $0.10 \pm 0.08$ & \citet{Torres2012} \\
\hline
\multirow{2}{1.9 cm}{\centering HD209458b} & \multirow{2}{1.8 cm}{\centering 0.69} & \multirow{2}{2 cm}{\centering \citet{Torres2008}} & $0.06-9.8^{\text{b}}$ & \citet{Line2016} & \multirow{2}{2.6 cm}{\centering $0.00 \pm 0.05$} & \multirow{2}{2 cm}{\centering \citet{Torres2008}} \\
 & & & $0.1-1.0^{\text{c}}$ & \citet{Brogi2017} & & \\
\hline
WASP-12b$^{\text{b}}$ & 1.41 & \citet{Hebb2009} & $0.3-20.0$ & \citet{Kreidberg2015b} & $0.07 \pm 0.07$ & \citet{Torres2012} \\
\hline
WASP-18b$^{\text{a}}$ & 10.2 & \citet{Triaud2010} & $0.6-2.0$ & \citet{Arcangeli2018} & $0.11 \pm 0.07$ & \citet{Torres2012} \\
\hline
WASP-39b$^{\text{b}}$ & 0.28 & \citet{Faedi2011} & $105-199$ & \citet{Wakeford2017b} & $-0.12 \pm 0.10$ & \citet{Faedi2011} \\
\hline
\multirow{2}{1.9 cm}{\centering WASP-43b} & \multirow{2}{1.8 cm}{\centering 2.03} & \multirow{2}{2 cm}{\centering \citet{Gillon2012}} & $0.4-1.7^{\text{b}}$ & \multirow{2}{2.6 cm}{\centering \citet{Stevenson2017}} & \multirow{2}{2.6 cm}{\centering $-0.05 \pm 0.17$} & \multirow{2}{2 cm}{\centering \citet{Hellier2011}} \\
& & & $0.3-1.7^{\text{d}}$ & & & \\
\hline
WASP-103b$^{\text{a}}$ & 1.49 & \citet{Gillon2014} & $10-53$ & Kreidberg et al., submitted & $0.06 \pm 0.13$ & \citet{Gillon2014} \\
\hline
\end{tabular}
\raggedright
$^{\text{a}}${From self-consistent modeling}

$^{\text{b}}${From H$_{2}$O detection}

$^{\text{c}}${From low-resolution + high-resolution spectroscopy}

$^{\text{d}}${From CO + CO$_{2}$}
\end{table*}

The 1D modeling also strongly constrains the dayside temperature because the atmosphere is almost isothermal over the bandpass observed by \textit{HST} and \textit{Spitzer}, and shows that HAT-P-7b has very poor heat redistribution. This agrees with a simple model of combined emitted and reflected light, which also shows that HAT-P-7b has a high dayside temperature, weak heat redistribution, and a low albedo. This is also in agreement with previous \textit{Spitzer} phase curves of HAT-P-7b, which showed that the planet had a hot dayside with $T=2667 \pm 57$\,K, a very low albedo, and weak heat redistribution \citep{Wong2016}. The weak heat redistribution predicted by these models fits the general trend that has been observed that planets at higher irradiation temperatures have less efficient heat redistribution, and therefore warmer daysides \citep{Schwartz2015}.

Although the 1D modeling indicates that HAT-P-7b has a thermal inversion due to absorption by TiO/VO, this inversion is not definitively observed because the data can be well-fit by a blackbody with $T=2692 \pm 14$\,K. The best-fitting 1D model suggests that the atmosphere does contain a thermal inversion, and that the WFC3 and \textit{Spitzer} data sample a part of the atmosphere near the tropopause where the T-P profile switches from non-inverted to inverted. In this region, the contribution functions are wide relative to the scale of the changes in the T-P profile, so the observations all appear to probe regions of similar temperatures and produce a featureless spectrum. However, this case can not be distinguished from a completely isothermal atmosphere. Even our model does not return a perfectly isothermal atmosphere, and it is possible that there are spectral features in the WFC3 bandpass below our level of precision. The WFC3 spectrum of WASP-18b, which has a signal-to-noise nearly four times that of our spectrum, shows some subtle emission features, and so a higher-precision spectrum of HAT-P-7b may reveal a similar structure \citep{Arcangeli2018}.

The 1D modeling also indicates that the blackbody-like spectrum of HAT-P-7b is produced because we are probing a range of pressures in the atmosphere, all of which have similar temperatures, and not just because only one pressure level is probed at all wavelengths. However, water dissociation in the upper atmosphere does limit the range of pressures probed. 
Other similar hot planets, like WASP-18b and WASP-103b, display muted spectral features in the WFC3 bandpass because of water dissociation and H$^{-}$ opacity \citep[Kreidberg et al. submitted, Parmentier et al. submitted]{Arcangeli2018}.

We also modeled the spectrum of HAT-P-7b with a 3D GCM. However, the GCM is unable to precisely reproduce the observed spectrum. Both the solar composition and the low-metallicity GCMs are colder on the dayside than the data suggest, which could be due to increased Lorentz forces causing more drag in the atmosphere or nightside clouds warming the planet. Additionally, the modeled spectra appear to have small emission features, which do not match the observed blackbody-like spectrum. In order to understand the cause of this discrepancy between the 3D GCM and the data for very hot planets like HAT-P-7b, more spectroscopic data are required over wider wavelength ranges and a larger parameter space exploration of parameters such as metallicity and Ti/O ratio in the GCM is needed.

Overall, the large dayside temperature of HAT-P-7b is extremely puzzling as the very low heat redistribution it implies cannot be reproduced by current GCMs. More observations and theoretical work are needed to understand what causes such a poor energy redistribution. Additionally, because our observations did not reveal any spectral features, further observations will be necessary to confirm the low value of the planet's metallicity. The \textit{James Webb Space Telescope} (\textit{JWST}) will have the ability to spectroscopically observe thermal emission of transiting planets over a large wavelength range. Spectroscopic observations of thermal emission with \textit{JWST} at the wavelengths covered by \textit{Spitzer}, where the GCM models for HAT-P-7b show the largest differences from each other, could resolve the discrepancy between the GCM and the observations. High-resolution spectroscopy \citep{Snellen2010} could also aid in understanding the thermal structure of HAT-P-7b by directly detecting the TiO responsible for the thermal inversion, as \citet{Nugroho2017} did for WASP-33b. Additionally, phase curves taken using \textit{JWST} or the \textit{Atmospheric Remote-sensing Exoplanet Large-survey} \citep[\textit{ARIEL},][]{Tinetti2016} could determine if nightside clouds are responsible for heating the dayside to temperatures above those predicted by the GCM.

\acknowledgements
Support for program GO-14792 was provided by NASA through a grant from the Space Telescope Science Institute, which is operated by the Association of Universities for Research in Astronomy, Inc., under NASA contract NAS 5-26555. J.L.B. acknowledges support from the David and Lucile Packard Foundation. M.R.L. acknowledges NASA XRP grant NNX17AB56G for partial support of the theoretical interpretation of the data as well as the ASU Research Computing staff for support with the Saguaro and Agave computing clusters. J.M.D. acknowledges that the research leading to these results has received funding from the European Research Council (ERC) under the European Union's Horizon 2020 research and innovation programme (grant agreement no. 679633; Exo-Atmos). D.D. acknowledges support provided by NASA through Hubble Fellowship grant HSTHF2-51372.001-A awarded by the Space Telescope Science Institute.


\end{document}